# THERMAL AND STRUCTURAL STABILITY OF MEDIUM ENERGY TARGET CARRIER ASSEMBLY FOR NOVA AT FERMILAB*

M.W. McGee[†], C. Ader, K. Anderson, J. Hylen, M. Martens
Fermi National Accelerator Laboratory, Batavia, IL 60510, USA

*Abstract*

The NOvA project will upgrade the existing Neutrino at Main Injector (NuMI) project beamline at Fermilab to accommodate beam power of 700 kW. The Medium Energy (ME) graphite target assembly is provided through an accord with the State Research Center of Russia Institute for High Energy Physics (IHEP) at Protvino, Russia. The effects of proton beam energy deposition within beamline components are considered as thermal stability of the target carrier assembly and alignment budget are critical operational issues. Results of finite element thermal and structural analysis involving the target carrier assembly is provided with detail regarding the target's beryllium windows.

## INTRODUCTION

The NuMI Low Energy (LE) target/baffle carrier design consisted of a stiff structural aluminum truss frame, which accommodated a longitudinal drive system for remote adjustment. The ME carrier design has been simplified, since the remote longitudinal drive system is no longer necessary. A preliminary analysis of the ME target and target casing completed by IHEP considered 120 GeV proton primary beam pulse with 1.3 mm (rms) spot size [1]. The carrier and target assembly will be exposed to a higher energy 700 kW beam heating load as compared to the 400 kW energy found in the NuMI beamline.

The ME target core is constructed from 50 ZXF-5Q grade graphite segments, 7.4 mm in width x 24.5 mm in length with a nominal overall target core length of 120 cm. Primary proton beam enters on the left of the target segments shown in Fig. 1 (colored red). An upstream (US) and downstream (DS) beryllium window hermetically seals the assembly within a helium environment to prevent oxidation of the graphite.

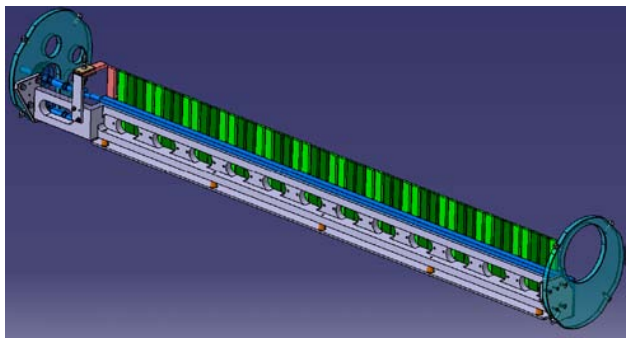

Figure 1: Solid model of ME graphite target assembly.

The baffle consists of ten ϕ57 mm O.D. x ϕ13 mm I.D. x 150 mm long graphite R7650 grade cores which are enclosed by a ϕ61 mm x 3 mm thick x 150 cm long aluminum tube after annealing. Eighteen 30-mm long radiator pin sections are evenly placed along its length. The graphite baffle prevents mis-steered primary proton beam from causing damage to the horn neck and target cooling/support components. It must withstand the full intensity of the beam for a few pulses during the time needed to detect the mis-steered beam and terminate beam. Also, the baffle is specified to be capable of continuous operation during 3% beam scraping at design luminosity.

The symmetric target carrier consists of two 30.5 cm aluminum c-channel x 1.46 m long sections which enclose the target assembly and graphite baffle as shown in Fig. 2. The twin c-channel design acts as fins to dissipate heat and provides structural stiffness (or stability) while allowing maximum access to the target and baffle.

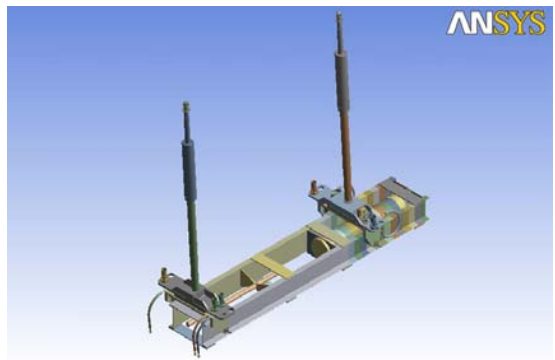

Figure 2: Solid model of ME target carrier assembly.

The target carrier assembly is supported by two 278.6 cm long shaft assemblies consisting of an upper ϕ146 mm 1018 cold-drawn steel shaft, a lower ϕ95.25 mm Invar steel shaft and a ϕ34.9 mm 1144 CF steel draw-bar. A heavy shielding module with positioning motors found above the carrier assembly provides support through these two shaft assemblies. The module provides motion control of the target, vertical and transverse to the beam by moving the shafts vertically and horizontally.

### Thermal Considerations

Beam heating occurs as showers of particles produced by proton beam interactions with the target pass through

---

the carrier, target casing, windows and DS hanger material. The carrier heating was considered within a MARS15 [2] simulation given the ME parameters shown in Table 1.

Table 1: NOνA (ME) beam parameters.

| Beam Energy | 120 GeV/c |
|---|---|
| Protons per Pulse | $4.90 \times 10^{13}$ |
| Cycle Time, sec | 1.33 |
| Beam Sigma, mm (rms) in (x,y) plane | 1.3 |
| Pulse Length, sec | $1 \times 10^{-6}$ |

Two time scales exist, the short 10 μsec pulse and the 1.33 second cool-off period found in between pulses. When considering the mass of a carrier, target casing and support structures, a steady-state analysis was applied in this case since the ΔT per pulse is rather small relative to the carrier structural members. Also, the 10 μsec spill time is a very rapid load in terms of thermal considerations. Almost no heat flow occurs on this time scale. Target carrier components are made of minimal volume-to-surface ratio aluminum and placed at the largest radius possible from the beamline. Beam heating found US of the target is negligible.

The beam deposits 9.9 kW of heating in the core and 4.7 kW in the target casing. Thermal radiation transfers 2.8 kW of core heating to the casing. The maximum temperature in the target core is 876 °C. Secondary particles also produce 0.2 kW of heat within the DS hanger [3]. These estimates consider a 700 kW beam with an additional 20% margin added to account for uncertainties of MARS15 calculations.

*Target Core, Casing and DS Hanger Cooling*

Two separate parallel cooling paths are considered; target core cooling and the DS hanger to target casing cooling serial path. Radioactive Water (RAW) flow, delivered at a temperature of $21 \pm 5$ °C removes 7.1 kW of energy from the target core and 7.7 kW of energy from the DS hanger and target casing.

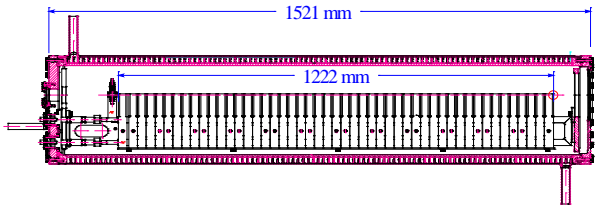

Figure 3: Graphite target within target casing assembly.

The target core cooling system consists of the graphite segments, cooling plates, pressing plates and connecting ϕ9 mm inner diameter tube as shown in Fig. 3. A target core cooling flow velocity of 2 m/sec corresponds to a water temperature rise of 12 °C. The target casing cooling system consists of a machined helically grooved ϕ268 mm x 11 mm thick wall aluminum pipe with a thin walled ϕ300 mm x 3 mm thick tube or welded jacket covering. RAW flow received DS through an inlet port at a velocity between 0.7 and 1.0 m/sec, exits US at roughly 26 °C with a 0.3 atm pressure drop. The 54 turn, 43 m long cooling channel path geometry is 22 mm width, 16.5 mm height and 6 mm fin thickness between channels with a calculated water heat transfer film coefficient of 4,903 (W/m$^2$/°C) [4].

## FINITE ELEMENT ANALYSIS

The finite element (FE) carrier assembly model generated using ANSYS Workbench calculates temperature profile and corresponding directional displacement due to thermal strain. In general for the carrier assembly, solid 70 and solid 90 (thermal elements) as well as solid 186 and solid 187 (structural elements) were used to construct the model.

Under a steady-state thermal response, convection was added as a surface condition with an air film coefficient value of 5 (W/m$^2$/°C) surrounding the carrier assembly. A conservative estimate of bulk air temperature surrounding the target pile during operation was 22 °C, based on NuMI operational thermocouple measurements from January 2009 [5].

Displacements of the carrier assembly due to thermal strain are calculated in the structural analysis. A cylindrical support was defined slightly above the module's bottom edge, which constrains each shaft radially. Relative joints between solid components were defined to allow for proper movement of the model as thermal strains increase. These included defining frictional connections at pivot points, based on component surfaces in contact. The US carrier pivot and DS pivot with linkage allows the carrier to find its natural position given the 3.34° beamline pitch.

*Thermal Results*

Under steady-state conditions, the target carrier shown in Fig. 4 reached a maximum temperature of 36.8 °C at the DS end. Target casing cooling played a limited role, since the six M8 x 1.25 support rods provide limited conduction. The DS hanger ears have water cooling, limiting vertical expansion.

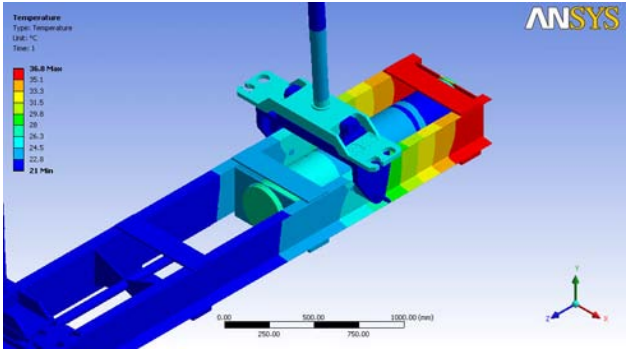

Figure 4: Temperature profile of target carrier.

*Displacement Results*

The calculated transverse (x) thermal displacement at beamline center was -1.23 x $10^{-2}$ mm (westerly), shown in Fig. 5. Fig. 6 depicts the vertical (y) thermal displacement of -0.193 mm (downward). Finally, Fig. 7 shows the longitudinal (z) thermal displacement of 2.92 x $10^{-2}$ mm (DS). All displacements were calculated with respect to the upper fixed supports.

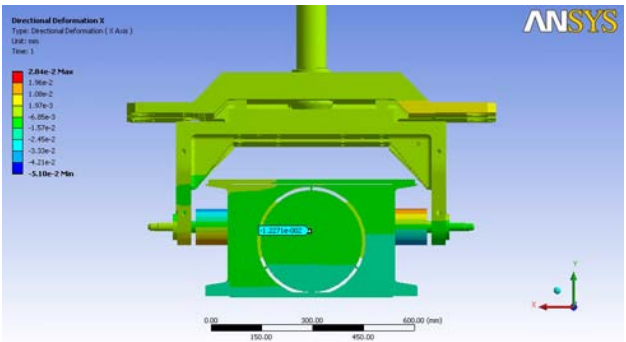

Figure 5: Transverse (x) displacement of target carrier.

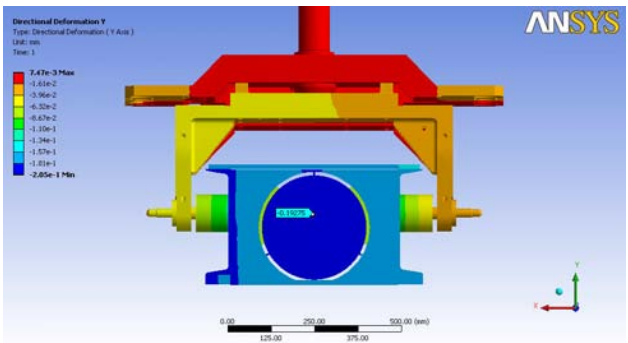

Figure 6: Vertical (y) displacement of target carrier.

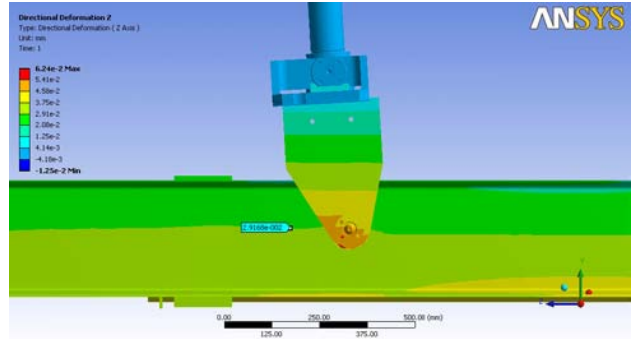

Figure 7: longitudinal (z) displacement of target carrier.

*Alignment Budget*

Misalignment of the proton beam, target and horn is a source of systematic error and therefore must be quantified. The target carrier assembly and the horn DS of it have essentially the same support system in terms of the heavy shielding module design and relative support within the target chase. The alignment budget allows for 0.25 mm of target carrier assembly thermal drift.

*US and DS Windows*

A transient thermal analysis was applied to each thin target window since the affect of rapid thermal pulsing could be seen. The US beryllium target window reaches thermal equilibrium after 30 seconds of beam or 23 pulses with a peak temperature of 66 $^{\circ}$C. The maximum von Mises stress (SEQV) of 223 MPa occurs at the edge of a 0.25 mm thick window when a 1.03 MPa helium load and ME deposition is applied.

Similarly, the DS beryllium target window reaches steady-state after 286 pulses with a peak temperature of 67.3 $^{\circ}$C. The maximum von Mises stress (SEQV) of 211 MPa occurs at the edge of a 1.25 mm thick window when a 1.03 MPa helium load and the beam heating deposition are applied. Both the US and DS window equivalent stresses were beneath the allowable stress of 224.1 MPa as defined by one half of the ultimate strength value [6] (with a safety factor of 2) and fatigue limit (considering >$10^7$ cycles) of 268 MPa.

## CONCLUSIONS

The overall thermal displacement of the carrier assembly from the vertical support rod connection at the top of module as it pertains to the thermal budget was investigated. Systematic alignment errors compound when considering the proton beam, target and horn relative position. The proposed requirement for support thermal drift is 0.25 mm due to heating from the beam. The total alignment budget for the target in the vertical and transverse direction considering all source terms is ±0.5 mm [7]. A westerly transverse movement of -1.23 x $10^{-2}$ mm, downward movement of -0.193 mm and 2.92 x $10^{-2}$ mm are calculated during beam operation. This analysis demonstrated that the ME carrier design meets the allowable thermal drift tolerance of 0.25 mm.

## ACKNOWLEDGEMENTS

We wish to thank Glenn Waver and Rod Stewart (CAD Support), Karl Williams (Target Water Group), Hiep Le and Keith Anderson (Target Group Support). Also, thanks to the Fermilab Alignment and Metrology Group (AMG) staff: Virgil Bocean, Gary Crutcher, Mike O'Boyle and Chuck Wilson. Special thanks to Fermilab computer support staff: Stewart Mitchell and Tony Metz.

## REFERENCES

[1] V. Garkusha et al., "Design Study of the NuMI Medium Energy Target for Higher Power Beams (Part II)," IHEP Report, (May 30, 2009).
[2] MARS Code System, http://www-ap.fnal.gov/MARS/.
[3] Private communication with Karl Williams (Fermilab).
[4] Private communication with Bryon Lundberg (Fermilab).
[5] Y. He and K. Anderson, "Thermal and Structural Analysis of Horn 1 Outer Conductor," Fermilab - MSDN-ME-000087, (February 24, 2008).
[6] J.L. Western, "Mechanical Safety Subcommittee Guidelines for Design of Thin Windows for Vacuum Vessels," Fermilab – TM-1380, (March 1993).
[7] M. Martens, J. Hylen and K. Anderson, "Target and Horn Configuration for the NOvA Experiment," Fermilab – NOVA-doc-3453, (November 9, 2009).